\documentclass[twocolumn,pra,aps,showpacs]{revtex4}         
\usepackage{graphicx}

\begin{document}

\title{Measurement of the interaction strength in a Bose-Fermi mixture with $^{87}$Rb and $^{40}$K}

\author{J. Goldwin, S. Inouye, M.L. Olsen, B. Newman\footnote{Current address: Department of Physics, MIT,
Cambridge, Massachusetts 02139, USA}, B.D.
DePaola\footnote{Permanent address: J. R. Macdonald Laboratory,
Department of Physics, Kansas State University, Manhattan, Kansas
66506, USA} and D.S. Jin\footnote{Quantum Physics Division,
National Institute of Standards and Technology.}}
\affiliation{JILA, National Institute of Standards and Technology,
University of Colorado, Boulder, Colorado 80309}
\date{11 August, 2004}

\begin{abstract}
A quantum degenerate, dilute gas mixture of bosonic and fermionic
atoms was produced using $^{87}$Rb and $^{40}$K. The onset of
degeneracy was confirmed by observing the spatial distribution of
the gases after time-of-flight expansion. Further, the magnitude
of the interspecies scattering length between the doubly spin
polarized states of $^{87}$Rb and $^{40}$K, $|a_{\rm RbK}|$, was
determined from cross-dimensional thermal relaxation. The
uncertainty in this collision measurement was greatly reduced by
taking the ratio of interspecies and intraspecies relaxation
rates, yielding $|a_{\rm RbK}| = 250 \pm 30\, a_0$, which is a
lower value than what was reported in [M. Modugno {\it et al.},
Phys. Rev. A {\bf 68}, 043626 (2003)]. Using the value for
$|a_{\rm RbK}|$ reported here, current $T=0$ theory would predict
a threshold for mechanical instability that is inconsistent with
the experimentally observed onset for sudden loss of fermions in
[G. Modugno {\it et al.}, Science {\bf 297}, 2240 (2002)].
\end{abstract}

 \maketitle

With many recent advances, ultracold Fermi gas experiments are now
beginning to explore the strongly interacting regime, where
pairing can lead to fermionic superfluidity \cite{BCSBEC}.  A
natural extension to such studies is to investigate the properties
of a mixture of a Fermi gas and a Bose-Einstein condensate (BEC).
The presence of a BEC can vastly modify the static and dynamic
character of a quantum degenerate Fermi gas. M\o lmer pointed out
that the mixture can phase separate, coexist, or collapse
depending on the relative sign and magnitude of the boson-boson
and boson-fermion scattering lengths \cite{molm98}. The spectrum
of collective excitations can show rich physics as the interaction
strength between the gases increases \cite{yip01}. In addition, an
effective fermion-fermion attraction arises from the boson-fermion
interaction, which can significantly increase $T_{\rm c}$ for a
phase transition to a BCS-type superfluid \cite{vive02}. This
mechanism for Cooper pairing of atoms is intriguing because of the
analogy to superconductivity.

To date, quantum degenerate atomic Bose-Fermi mixtures have been
realized in the following combinations: $^{6}$Li (fermion) with
$^{7}$Li (boson) \cite{LiLi}, $^{6}$Li with $^{23}$Na (boson)
\cite{hadz02}, and $^{40}$K (fermion) with $^{87}$Rb (boson)
\cite{roat02}. With the last mixture, pioneering experiments by
the LENS group revealed that the scattering length between
$^{87}$Rb and $^{40}$K is large and negative
\cite{modu02,roat02,ferr02,modu03}. At high densities, this strong
attractive interaction can exceed the Pauli pressure and cause the
whole system to become mechanically unstable and collapse
\cite{molm98,roth02,rothonly02,miya01,modu03}. The LENS group has
reported the observation of such a collapse; they see a sudden
loss of $^{40}$K atoms as the number of atoms in the $^{87}$Rb BEC
exceeds a threshold value \cite{modu02}.

\begin{figure}\includegraphics[scale=0.7]{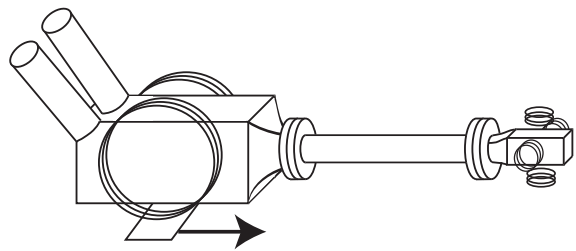}
\caption{Schematic of the experimental apparatus. The glass cell
on the left (``MOT cell'') is filled with a background vapor of
$^{87}$Rb and $^{40}$K released from alkali-metal dispensers
attached to the left end of the cell. The cell on the right
(``science cell'') is under ultrahigh vacuum, and small coils
surrounding the cell produce a magnetic trap with strong
confinement for evaporative cooling. (The figure does not show all
the coils.) Atoms are transferred from the MOT cell to the science
cell using a moving magnetic trap produced by an anti-Helmholz
coil pair on a translation stage. \label{fig:schematic}}
\end{figure}

Here we report the production of a quantum degenerate Bose-Fermi
mixture with $^{87}$Rb and $^{40}$K atoms and a determination of
the magnitude of the cross-species scattering length. Our
experimental apparatus for producing the mixture, which requires
only a single magneto-optical trap (MOT), is outlined in Fig.
\ref{fig:schematic}.  In brief, our vacuum chamber consists of two
glass cells --- one contains a two-species MOT while the other
cell (science cell) is used for evaporative cooling of the
mixture. The lifetime for $^{87}$Rb atoms in the magnetic trap,
which is limited by collisions with the background gas, is $\sim
4$ s in the MOT cell and more than 100 s in the science cell. The
atoms are transferred between the cells using a moving magnetic
trap.

The experimental procedure for creating the mixture is as follows.
First, $2\times 10^9$ $^{87}$Rb atoms and $1\times 10^7$ $^{40}$K
atoms are collected in a two-species MOT from the background vapor
\cite{gold02}. The atoms are then transferred into a quadrupole
magnetic trap, whose magnetic field gradient along the symmetry
axis in the horizontal direction is 99 G/cm. During the last 50 ms
of the MOT, the $^{87}$Rb cloud is compressed by reducing the
intensity of the hyperfine repumping light. Simultaneously the
$^{40}$K cloud is Doppler-cooled by reducing the detuning of the
trapping light to about one natural linewidth ($\sim 6 $ MHz). In
addition, the atoms are optically pumped to their doubly spin
polarized states, which are $|F,m_F\rangle = |2,2\rangle$ for
$^{87}$Rb and $|9/2, 9/2\rangle$ for $^{40}$K. Once in the
quadrupole magnetic trap, the atoms are transferred to the science
cell by physically moving the trapping coils, which are mounted on
a motorized translation stage \cite{lewa02}. The observed loss in
the phase-space density of the $^{87}$Rb cloud during the 81 cm (6
sec) travel is only a factor of two. In the science cell, the
atoms are transferred to an Ioffe-Pritchard-type magnetic trap for
the final cooling. Since there is no MOT in the science cell, the
cell size can be small (our cell is 1 ${\rm cm}^2$ in
cross-section) and we can place coils very close to the atoms to
achieve tight confinement with relatively low electric current
($\sim$ 27 A). The radial (axial) trapping frequency for $^{87}$Rb
is $\nu_{r,{\rm Rb}}=$ 160 Hz ($\nu_{z,{\rm Rb}}=$ 25 Hz), and the
trapping frequencies for $^{40}$K are $\sqrt{m_{\rm Rb}/m_{\rm
K}}=1.47$ times higher, where $m_{\rm Rb}$ ($m_{K}$) is the mass
of $^{87}$Rb ($^{40}$K) atoms.

Simultaneous quantum degeneracy is achieved by radio frequency
(rf)-induced evaporative cooling of $^{87}$Rb and sympathetic
cooling of $^{40}$K. This scheme circumvents the problem of
vanishing elastic cross section for spin polarized fermions at
ultracold temperatures \cite{dema99}. The typical initial number
and temperature of the $^{87}$Rb atoms in the Ioffe-Prichard trap
before evaporation are $N_{\rm Rb} = 3 \times 10^8$ and $T_{\rm
Rb} = 400\; \mu {\rm K}$; with these initial conditions, we
produce a nearly pure BEC of $2 \times 10^5$ $^{87}$Rb atoms after
40 seconds of evaporation. The initial conditions for $^{40}$K in
the Ioffe-Prichard trap are not known due to the difficulty of
measuring a small optical density cloud with absorption imaging.
For sympathetic cooling from 10 $\mu{\rm K}$ to 0.4 $\mu{\rm K}$,
a range where the $^{40}$K cloud can be imaged and there is not
yet a $^{87}$Rb BEC, we observe less than 10\% number loss with
typically $1.4 \times 10^5$ $^{40}$K atoms.

\begin{figure}\includegraphics[scale=0.7]{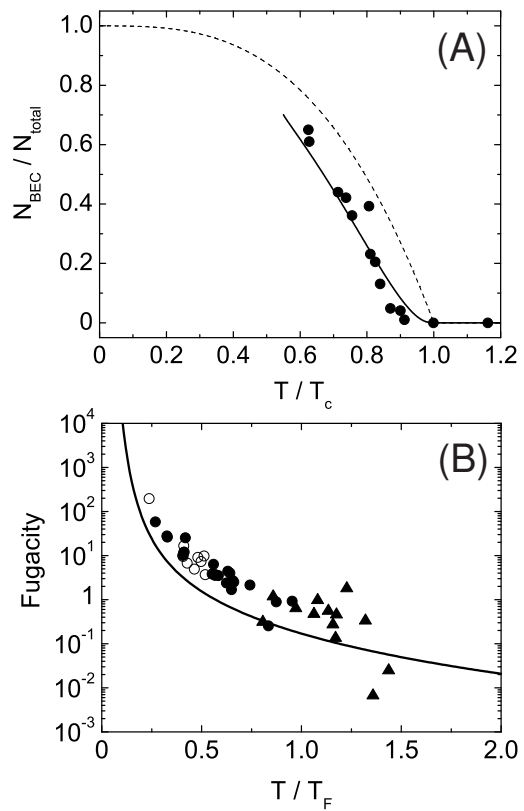}
\caption{Quantum degeneracy observed in time-of-flight images. (A)
Condensate fraction $N_{\rm BEC}/N_{\rm total}$ obtained from 20
ms time-of-flight images of $^{87}$Rb. The dashed line shows the
condensate fraction for an ideal Bose gas in the thermodynamic
limit. The solid line includes the effects from the finite number
of atoms and the positive chemical potential due to boson-boson
interactions in the condensate \cite{nara98}. The phase transition
occurs at 440 nK. (B) Fugacity of the fermionic cloud obtained by
fitting the density distribution of $^{40}$K atoms after 5 ms
time-of-flight. The fugacity $z$ is equal to the peak phase space
density of the gas in the classical limit. The different symbols
denote data taken with a $^{87}$Rb BEC present (filled circles),
with a $^{87}$Rb thermal cloud present (filled triangles), and
without $^{87}$Rb atoms present (open circles). The data are
systematically shifted from the theory curve expected for an ideal
Fermi gas (line). A typical value of $T_F$ at $T/T_F$ = 0.5 is 220
$\pm$ 50 nK. \label{fig:degeneracy} }
\end{figure}

Quantum degeneracy for the $^{87}$Rb gas is evident in the
formation of a Bose-Einstein condensate, which is characterized by
the emergence of a bimodal distribution in the absorption images
taken after time-of-flight expansion. The fraction of atoms in the
BEC is determined by fitting the distribution with the sum of the
Thomas-Fermi profile (inverted parabola) for the condensate and
the thermal distribution for an ideal Bose gas (Fig.
\ref{fig:degeneracy}(A)). The temperature is determined from the
fit to the noncondensed component. For comparison to our data, a
theory curve that includes both the effects from finite number and
the positive chemical potential due to boson-boson interactions in
the condensate is also shown \cite{nara98}. The theory and the
experiment are in good agreement. These data were taken with $\sim
1 \times 10^4$ $^{40}$K atoms present at the end of evaporation.
In comparing data taken with and without $^{40}$K atoms present,
no effect of $^{40}$K on the $^{87}$Rb condensate fraction is
observed. This is consistent with the low $^{40}$K density($\sim 3
\times 10^{12}\, {\rm cm}^{-3}$), which results in a negligible
contribution to the $^{87}$Rb mean field.

The effect of quantum degeneracy on fermions is more subtle. The
density distribution in time-of-flight becomes more
``flat-topped'' than that for a classical gas due to the Pauli
pressure. The fugacity $z = \exp(\mu_K/k_B T)$ (where $\mu_{\rm
K}$ is the chemical potential of $^{40}$K), the cloud size in each
direction, and the peak density of the gas are used as independent
parameters for fitting the images. In Fig. \ref{fig:degeneracy}(B)
the measured fugacity is plotted as a function of $T/T_{F}$, where
$T$ is determined from the cloud size, and $T_{F}$ is the Fermi
temperature. From the fugacity measurement and assuming an ideal
Fermi gas, we find that the lowest $T/T_F$ achieved in this
experiment was 0.20. For comparison to the data, the solid line
shows the prediction for an ideal Fermi gas in a 3D harmonic trap
\cite{butt97}. Most of the data lie to the right of the curve.
This discrepancy could be explained if our $^{40}$K number
calibration is low by about a factor of two so that $T/T_F$ is
systematically overestimated by 20\%. Note that the fugacity,
which is a measure of the ``flatness'', is free from calibration
errors. The temperature measurement has been confirmed by fitting
to the thermal wing of $^{87}$Rb images taken during the same
experimental run. Finally, we note that the measured fugacity
shows no systematic difference between $^{40}$K images taken with
and without a $^{87}$Rb BEC present in the trap. With our highest
number of $1.8 \times 10^5$ $^{87}$Rb atoms in a BEC and $8 \times
10^4$ $^{40}$K atoms in the Fermi gas we observe only a slow
number loss ($\sim 1$ sec timescale) and no sudden collapse.

The strength of inter-species interactions is critical in
determining the static and dynamic properties of the mixture. We
have determined the magnitude of the interspecies scattering
length $|a_{\rm RbK}|$ through measurements of cross-dimensional
thermal relaxation \cite{monr93}. In general, this method has
limited accuracy because of systematic uncertainties in the atom
number, which is typically determined from absorption imaging.
Experimental imperfections such as frequency jitter, imperfect
polarization of the probe beam, or fluctuation and tilt of the
magnetic field during imaging tend to reduce probe beam
absorption. This leads to an underestimate of the number of atoms
and an overestimate of the cross section. Here we avoid this
systematic error by taking the ratio of two cross section
measurements that are both proportional to the number of $^{87}$Rb
atoms: (1) thermal relaxation of $^{40}$K atoms in the presence of
$^{87}$Rb and (2) relaxation of a single-species $^{87}$Rb cloud.
We further use the fact that the Rb-Rb cross section is already
known with high accuracy \cite{kemp02}.

The experimental procedure is as follows. First, we prepare
$1\times 10^5$ $^{40}$K atoms and $4 \times 10^5$ $^{87}$Rb atoms
at temperatures $T\sim 1\mu$K in a magnetic trap whose radial
trapping frequency $\nu_{r, {\rm Rb}}$ is 90 Hz. Both gases are
away from quantum degeneracy and elastic collisions are in the
$s$-wave limit. Multiple 1 ms rf sweeps can be applied to reduce
the number of $^{87}$Rb atoms by as much as a factor of four, for
the purpose of varying the density. The bias magnetic field is
then ramped from 4.3 G to 1.4 G to increase the radial confinement
of the magnetic trap to its regular strength ($\nu_{r,{\rm Rb}}=$
160 Hz); this adds energy to the radial direction. The duration of
the ramp is kept long ($\ge 10$ ms) compared to the radial
trapping period to avoid excitation of collective modes, but short
compared to the relaxation time in the gas. This is always
possible unless the gases enter into the hydrodynamic regime.

\begin{figure}\includegraphics[scale=0.9]{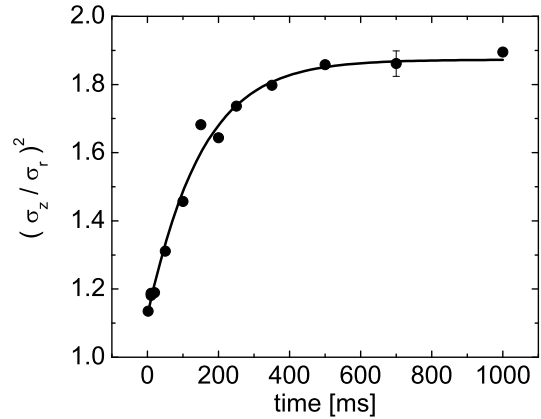}
\caption{Typical relaxation curve for the energy anisotropy. Shown
is the the square of the $^{40}$K cloud aspect ratio after 5 ms
time-of-flight; this is proportional to the ratio of the energies
in each direction. The $^{40}$K cloud rethermalizes through
elastic collisions with $^{87}$Rb atoms. The line is a fit to the
ratio of exponentially decaying curves. The error bar represents
only the statistical uncertainty. \label{fig:exponential}}
\end{figure}

Just after the ramp, the energy of the gas in the radial direction
is higher than in the axial direction. The cross-dimensional
rethermalization time constant is extracted by holding the mixture
for a variable time in the final trap and measuring the aspect
ratio after time-of-flight. Shown in Fig. \ref{fig:exponential} is
a typical relaxation curve for the energy anisotropy. Rather than
the radial and axial sizes, the aspect ratio of the cloud,
$\sigma_z/\sigma_r$, is used for data analysis since it is less
sensitive to shot-to-shot variations in the absolute temperature
of the cloud before the compression \cite{asymptote}.

The data analysis is simplified by two facts: (i) due to the Fermi
statistics, the spin-polarized $^{40}$K atoms collide only with
$^{87}$Rb atoms, and (ii) in our experiment there is no net flow
of energy between the $^{40}$K and $^{87}$Rb gases. Therefore, the
thermal relaxation rate of the $^{40}$K cloud, $1/\tau_{\rm K}$,
depends only on the collision rate with $^{87}$Rb atoms
\cite{bloc01}:
\begin{equation}
\frac{1}{\tau_{\rm K}}= \frac{1}{\beta}\,\langle n_{\rm Rb}
\rangle\, \sigma_{\rm Rb K}\, v_{\rm Rb K} \; {}_{.}
\label{tauKRb}
\end{equation}
Here, $\langle n_{\rm Rb}\rangle$ is the average density of
$^{87}$Rb atoms in the trap, $\sigma_{\rm Rb K}=4 \pi a_{\rm Rb
K}^2$ is the elastic cross section for Rb-K collisions, and
$v_{\rm Rb K}= \sqrt{8 k_B T / \pi \mu}$ is the thermally averaged
relative speed between $^{87}$Rb and $^{40}$K atoms, where $\mu =
m_{\rm Rb} m_{\rm K}/(m_{\rm Rb} +m_{\rm K})$ is the reduced mass.
The coefficient of proportionality, $\beta$, can be regarded as
the number of collisions per $^{40}$K atom needed for thermal
relaxation. The relaxation rate for a single-species $^{87}$Rb
cloud $1/\tau_{\rm Rb}$ is proportional to the Rb-Rb collision
rate:
\begin{equation}
\frac{1}{\tau_{\rm Rb}}= \frac{1}{\alpha}\,\langle n_{\rm Rb}
\rangle \, \sigma_{\rm Rb Rb}\, v_{\rm Rb} \; {}_, \label{tauRb}
\end{equation}
where $\sigma_{\rm Rb Rb}=8 \pi a_{\rm Rb}^2$ is the elastic cross
section for Rb-Rb collisions, $v_{\rm Rb}= 4 \sqrt{k_B T / \pi
m_{\rm Rb}}$ is the thermally averaged relative speed between
$^{87}$Rb atoms, and $\alpha$ is the coefficient of
proportionality for single-species relaxation. Thus, the ratio of
the two time constants is free from systematic uncertainties in
the $^{87}$Rb number \cite{overlap}.

\begin{equation}
\frac{\tau_{\rm Rb}}{\tau_{\rm K
}}=\frac{1}{2}\,\frac{\alpha}{\beta}\, \sqrt{\frac{m_{\rm
Rb}}{2\mu}}\, \left(\frac{a_{\rm Rb K}}{a_{\rm Rb}} \right)^2 \;
{}_{.} \label{ratio}
\end{equation}

The constant $\alpha$ ($\beta$) has been determined to be $2.67
\pm 0.13$ ($2.1 \pm 0.2$) from Monte Carlo simulations using our
experimental parameters, including the initial energy anisotropy
of both species and the mass difference between $^{40}$K and
$^{87}$Rb \cite{gold04}.  It is this mass difference that gives
rise to the difference between $\alpha$ and $\beta$, as a lighter
particle can redistribute more of its total kinetic energy in a
single collision.

\begin{figure}\includegraphics[scale=0.85]{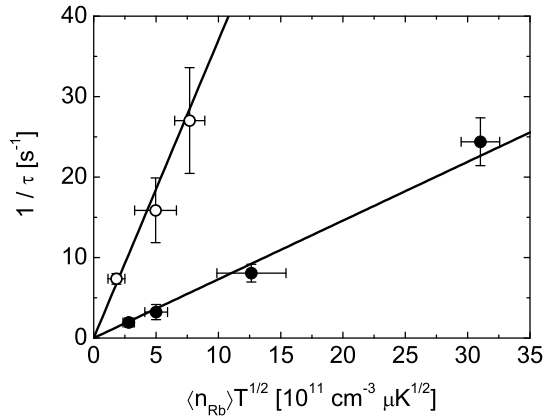}
\caption{Measurement of elastic collision cross-section through
energy anisotropy relaxation. Relaxation time constants for a
$^{40}$K gas in the presence of $^{87}$Rb (open circles) and
$^{87}$Rb gas alone (filled circles) are shown as a function of
the product of average $^{87}$Rb density in the trap and the
square root of the equilibrium temperature. Lines are linear fits
through the origin. The ratio of the slopes is used to extract the
magnitude of interspecies scattering length $|a_{\rm Rb K}|$, as
shown in Eq. (\ref{ratio}).}\label{fig:slopes}
\end{figure}

The measured relaxation rates for varying density of $^{87}$Rb
atoms are plotted in Fig. \ref{fig:slopes}. The error bars only
account for statistical errors. Systematic uncertainties in the
$^{87}$Rb density were eliminated from the determination of
$|a_{\rm RbK}|$ by taking the ratio of the two slopes and using
Eq. (\ref{ratio}). The known value $a_{\rm Rb}=98.98 \pm 0.04\,
a_0 $ for $^{87}$Rb in the $|2,2\rangle$ state \cite{kemp02},
obtained by combining multiple high-precision measurements
including photoassociation and Feshbach spectroscopy, is used
 for calibration \cite{aRb} and we obtain $|a_{\rm RbK}| = 250
\pm 30\, a_0$.

This measured value of $|a_{\rm RbK}|$ is in mild disagreement
with the value of $-410^{+80}_{-90}\, a_0$ from the LENS group
\cite{modu02}, which was determined from a collisional measurement
that had the usual large systematic number uncertainty.  More
serious disagreement is seen when comparing our result with the
value of $-395\pm 15\, a_0$ determined from a comparison of theory
predictions and experimental observation of collapse phenomena
\cite{modu03}. These results suggest a need for further
investigation of the behavior of the mixture close to the
collapse. For example the zero temperature phase diagram in ref.
\cite{rothonly02} which assumed $a_{\rm RbK} = -260\, a_0$
predicts that a condensate with $2\times 10^6$ atoms remains
mechanically stable with numbers of $^{40}$K atoms as high as
$1\times 10^7$. The new value for $|a_{\rm RbK}|$ that we report
here is also likely to have an impact on the predicted positions
of Feshbach resonances \cite{simo03,BFFRs}.

\begin{acknowledgments}
The authors thank Eric Cornell, and Carl Wieman for useful
discussions. We acknowledge support from the U.S. Department of
Energy, Office of Basic Energy Sciences, the W. M. Keck
Foundation, and the National Science Foundation. B.D.D. expresses
his thanks to the JILA Visiting Fellow program.
\end{acknowledgments}

\bibliographystyle{prsty}

\begin{thebibliography}{99}
\bibitem{asymptote}
The measured equilibrium aspect ratio is higher than that for an
isotropic energy distribution with purely ballistic expansion.  We
attribute this to imperfections in the trap turn-off, and include
a 10\% uncertainty in $T$ in our error bar for $|a_{\rm RbK}|$.
\bibitem{overlap} Note that the average density of $^{87}$Rb ($\langle n_{\rm
Rb}\rangle$) in Eq. (\ref{tauKRb}) is taken over the $^{40}$K
density distribution, while that in Eq. (\ref{tauRb}) is taken
over the $^{87}$Rb distribution. These differ only by a
temperature-dependent factor due to the different equilibrium
positions of the clouds due to gravity. For these data this effect
is limited to 3 to 5\%, and is included in the analysis.
\bibitem{aRb} Eq.(\ref{tauRb}) with
single-species data in Fig. \ref{fig:slopes} yields $a_{\rm
Rb}=109\, a_0$, which is 10\% larger than the accepted scattering
length of $^{87}$Rb in the $|F,m_F\rangle=|2,2\rangle$ states.
This corresponds to a 20\% underestimation of our $^{87}$Rb
number.
\end{thebibliography}


\end{document}